\documentclass[12pt,preprint2]{emulateapj}

\usepackage{upgreek} 
\usepackage{gensymb} 
\usepackage{xcolor}

\usepackage{graphicx,epstopdf}

 \newcommand\omicron{o}

\slugcomment{Draft v17}

\shorttitle{Deep LBTI observations of $\beta$~Leo}
\shortauthors{Defr\`ere et al.}

\begin{document}

\title{The HOSTS survey: evidence for an extended dust disk and constraints on the presence of giant planets in the Habitable Zone of $\beta$~Leo} 

\author{D.~Defr\`ere\altaffilmark{1,2},
P.M.~Hinz\altaffilmark{3},
G.M.~Kennedy\altaffilmark{4}, 
J.~Stone\altaffilmark{5,$\star$}, 
J.~Rigley\altaffilmark{6}, 
S.~Ertel\altaffilmark{7,8}, 
A.~Gaspar\altaffilmark{7}, 
V.P.~Bailey\altaffilmark{9},
W.F.~Hoffmann\altaffilmark{7},  
B.~Mennesson\altaffilmark{9}, 
R.~Millan-Gabet\altaffilmark{10},  
W.C.~Danchi\altaffilmark{11},
O.~Absil\altaffilmark{2,$\star\star$}, 
P.~Arbo\altaffilmark{7}, 
C~ Beichman\altaffilmark{10}, 
M~ Bonavita\altaffilmark{12,13}, 
G.~Brusa\altaffilmark{7}, 
G.~Bryden\altaffilmark{9},  
E.C.~Downey\altaffilmark{7}, 
S.~Esposito\altaffilmark{14}, 
P.~Grenz\altaffilmark{7}, 
C.~Haniff\altaffilmark{15}, 
J.M.~Hill\altaffilmark{8}, 
J.M.~Leisenring\altaffilmark{7}, 
J.R.~Males\altaffilmark{7}, 
T.J.~McMahon\altaffilmark{7}, 
M.~Montoya\altaffilmark{7}, 
K.M.~Morzinski\altaffilmark{7}, 
E.~Pinna\altaffilmark{14}, 
A.~Puglisi\altaffilmark{14}, 
G.~Rieke\altaffilmark{7}, 
A.~Roberge\altaffilmark{11}, 
H.~Rousseau\altaffilmark{2}, 
E.~Serabyn\altaffilmark{9}, 
E.~Spalding\altaffilmark{7,16}, 
A.J.~Skemer\altaffilmark{3}, 
K.~Stapelfeldt\altaffilmark{9}, 
K.~Su\altaffilmark{7},  
A.~Vaz\altaffilmark{7}, 
A.J.~Weinberger\altaffilmark{17},
M.C.~Wyatt\altaffilmark{6} }

\affil{\altaffilmark{1}Institute of Astronomy, KU Leuven, Celestijnenlaan 200D, 3001, Leuven, Belgium}
\affil{\altaffilmark{2}Space sciences, Technologies \& Astrophysics Research (STAR) Institute, University of Li\`ege, Li\`ege, Belgium}
\affil{\altaffilmark{3}University of California, Santa Cruz, Santa Cruz, 1156 High Street, Santa Cruz, CA 95064, USA}
\affil{\altaffilmark{4}Department of Physics, University of Warwick, Gibbet Hill Road, Coventry CV4 7AL, UK}
\affil{\altaffilmark{5}Naval Research Laboratory, Remote Sensing Division, 4555 Overlook Ave SW, Washington, DC 20375}
\affil{\altaffilmark{6}Institute of Astronomy, University of Cambridge, Madingley Road, Cambridge CB3 0HA, UK}
\affil{\altaffilmark{7}Steward Observatory, Department of Astronomy, University of Arizona, 933 N. Cherry Ave, Tucson, AZ 85721, USA}
\affil{\altaffilmark{8}Large Binocular Telescope Observatory, University of Arizona, 933 N. Cherry Ave, Tucson, AZ 85721, USA}
\affil{\altaffilmark{9}Jet Propulsion Laboratory, California Institute of Technology, 4800 Oak Grove Drive, Pasadena CA 91109-8099, USA}
\affil{\altaffilmark{10}California Institute of Technology, NASA Exoplanet Science Institute, Pasadena, CA 91125, USA}
\affil{\altaffilmark{11}NASA Goddard Space Flight Center, Exoplanets \& Stellar Astrophysics Laboratory, Code 667, Greenbelt, MD 20771, USA}
\affil{\altaffilmark{12}SUPA, Institute for Astronomy, University of Edinburgh, Blackford Hill, Edinburgh EH9 3HJ, UK}
\affil{\altaffilmark{13}Centre for Exoplanet Science, University of Edinburgh, Edinburgh EH9 3HJ, UKA}
\affil{\altaffilmark{14}INAF-Osservatorio Astrofisico di Arcetri, Largo E. Fermi 5, I-50125 Firenze, Italy}
\affil{\altaffilmark{15}Cavendish Laboratory, University of Cambridge, JJ Thomson Avenue, Cambridge CB3 0HE, UK}
\affil{\altaffilmark{16}Department of Physics, University of Notre Dame, 225 Nieuwland Science Hall, Notre Dame, IN, 46556, USA}
\affil{\altaffilmark{17}Department of Terrestrial Magnetism, Carnegie Institution of Washington, 5241 Broad Branch Road NW, Washington, DC, 20015, USA}

\email{$\star$ Hubble Fellow.}
\email{$\star\star$ FNRS Research Associate.}

\begin{abstract}
The young (50--400\,Myr) A3V star $\beta$~Leo is a primary target to study the formation history and evolution of extrasolar planetary systems as one of the few stars with known hot ($\sim$1600$\degree$K), warm ($\sim$600$\degree$K), and cold ($\sim$120$\degree$K) dust belt components. In this paper, we present deep mid-infrared measurements of the warm dust brightness obtained with the Large Binocular Telescope Interferometer (LBTI) as part of its exozodiacal dust survey (HOSTS). The measured excess is 0.47\%$\pm$0.050\% within the central 1.5\,au, rising to  0.81\%$\pm$0.026\% within 4.5\,au, outside the habitable zone of $\beta$~Leo. This dust level is 50 $\pm$ 10 times greater than in the solar system's zodiacal cloud. Poynting-Robertson drag on the cold dust detected by Spitzer and Herschel under-predicts the dust present in the habitable zone of $\beta$~Leo, suggesting an additional delivery mechanism (e.g.,~comets) or an additional belt at $\sim$5.5\,au. A model of these dust components is provided which implies the absence of planets more than a few Saturn masses between $\sim$5~au and the outer belt at $\sim$40~au. We also observationally constrain giant planets with the LBTI imaging channel at 3.8~$\upmu$m wavelength. Assuming an age of 50\,Myr, any planet in the system between approximately 5\,au to 50\,au must be less than a few Jupiter masses, consistent with our dust model. Taken together, these observations showcase the deep contrasts and detection capabilities attainable by the LBTI for both warm exozodiacal dust and giant exoplanets in or near the habitable zone of nearby stars.
\end{abstract}


\section{Introduction}
\label{sec:intro}

Debris material surrounding mature stars provides important insight into the architecture and potential formation history of a star's planetary system. Removal mechanisms such as collisional cascades \citep[e.g.,][]{Gaspar2012,Kenyon2016}, leading to the blowout of small grains by radiation pressure, or Poynting-Robertson (P-R) drag, clear the dust grains out over short timescales, relative to the lifetime of the star \citep[e.g.,][]{Wyatt2008}. Thus, debris material is an indication of solid planetesimals orbiting the star. In our own solar system, debris material is created by parent bodies from both the asteroid belt, at 3-4\,au, and the Kuiper belt at 40\,au. While most of this debris dust orbits near to where it is produced, some is delivered to Earth's vicinity, where it is detected as the zodiacal dust. The relative contribution from each parent body has been refined by dynamical analysis of the dust distribution. Previous estimates have indicated the dust is dominated by material from the asteroid belt \citep{Dermott2002}, while a more recent analysis has favored the source material being from comets interacting with Jupiter and the other giant planets \citep{Nesvorny2010}.

Debris dust in the habitable zones of other stars - typically referred to as exozodiacal dust - is likewise thought to come from extrasolar asteroids and comets. This is supported by observations from the Keck Interferometer Nuller (KIN), which find exozodiacal dust primarily around stars with known (Kuiper-Belt-like) dust belts in the outer regions \citep{Mennesson2014}. A simple analytic model of P-R drag and collisions \citep{Wyatt2005} is consistent with these results. However, more recent studies find that this model probably under-predicts the dust level in the habitable zone (HZ) by a factor of at least several \citep{vanLieshout2014,Kennedy2015b,Rigley2020}, which suggests that the models require further work, and/or that the HZ dust detected by the KIN has a significant contribution from processes other than P-R drag \citep[e.g., comets][]{Faramaz2017}. Recently, this trend has been supported by new observations from the Hunt for Observable Signatures of Terrestrial planetary Systems \citep[HOSTS,][]{Danchi2016,Ertel2018,Ertel2020} on the Large Binocular Telescope Interferometer \citep[LBTI,][]{Hinz2003,Hinz2004,Hinz2016,Ertel2020b}. In particular, \cite{Ertel2020} report that 10 of the 38 stars observed show significant excess and a clear correlation with the presence of cold dust in the systems but no correlation with the spectral type of the host star. Following the detailed analysis of the star with the highest infrared excess \citep[i.e., $\eta$~Crv,][]{Defrere2015}, we model in this paper the data obtained with the LBTI on another particularly interesting star of the HOSTS survey: $\beta$~Leo (Denebola, HD\,102647, HIP\,57632). As one of the brightest stars of the HOSTS target list, and due to its scientific value (see below), $\beta$~Leo was one of the first stars observed during the survey. Located at a distance of 11.0\,pc, this A3V star is a primary target to study the formation history and evolution of extrasolar planetary systems as one of the few stars with known hot ($\sim$1600$\degree$K), warm ($\sim$600$\degree$K), and ($\sim$120$\degree$K) cold dust belt components \citep{Churcher2011}. The debris disk was first detected by unresolved  space-based photometric observations using IRAS \citep{Aumann1985} and Spitzer \citep{Su2006,Chen2006}. Spatially resolved observations at 100~$\upmu$m and 160~$\upmu$m with Herschel resolved a cold disk extending to $\sim$40\,au \citep{Matthews2010} with a position angle of 125$\degree\pm$15$\degree$ and an inclination of 57$\degree\pm$7$\degree$ from edge on \citep{Churcher2011}. The first spatially-resolved observations at mid-infrared wavelengths were obtained with the MMT, leading \citet{Stock2010} to propose a two-component dust model with planetesimal belts at 2 -- 3\,au and 5 -- 55\,au.  \citet{Churcher2011} suggests a three-component model with blackbody belts at 2\,au, 9\,au, and 30-70\,au. Further spatially-resolved observations at mid-infrared wavelengths with the KIN concluded that the inner belt must reside between 0.07 and 2.2\,au \citep{Mennesson2014}. $\beta$~Leo is actually one of the rare systems with a spatially resolved warm dust distribution, a phenomenon mostly observed with mid-infrared interferometry \citep[e.g.,][]{Mennesson2013,Defrere2015,Lebreton2016} and, in a few favorable cases, with single-dish 8-m class imaging telescopes \citep{Moerchen2007,Moerchen2010,Smith2009}.

As a nearby young star, $\beta$~Leo is a prime target for direct imaging campaigns, but no exoplanet detections have been reported so far \citep{Meshkat2015,Durkan2016}. $\beta$~Leo is a relatively young A-star, for which isochronal age estimates are generally very unreliable. The isochrones age study by \cite{Nielsen2013} finds 95\% probability limits for the age of  16 – 45\,Myr, while a similar study by \cite{David2015} assigns limits of 61 – 649\,Myr. Other assignments of age from isochrones populate these ranges. However, it is believed that the star is a member of the Argus moving group \citep{Zuckerman2011,Zuckerman2019,Baron2019}. The age of this moving group can be determined relatively accurately from isochrones, lithium abundances, etc. for the lower-mass members and is believed to be 55 – 70 Myr \citep[with uncertainties due to possible contamination,][]{Bell2015},  40 – 50\,Myr \citep{Zuckerman2019}, or 30 – 40\,Myr \citep{Lee2019}. If this membership assignment is correct, then a rough lower limit to the age is 50\,Myr. If instead $\beta$~Leo is a field star, then isochrones are the only way to assign an age. A rough upper limit is 412\,Myr \citep{Stone2018}. This value was computed by incorporating knowledge of the local stellar population to implement a Bayesian approach to derive a posterior distribution function of age, mass, and the metallicity ratio with respect to the Sun. To capture the uncertainties, we have used both of these limits in our analysis.



The goal of this paper is to present a general picture of the $\beta$~Leo system that is consistent with both our new LBTI observations and ancillary data found in the literature. We present the instrumental setup and configuration of the LBTI in section \ref{sec:instr}. In section \ref{sec:obs}, we describe the observations obtained with both the nulling instrument and the L-band direct imaging camera. Section \ref{sec:reduction} then gives an overview of the data reduction approach for each observation and Section~\ref{sec:results} describes the results. We conclude this paper by a discussion (Section \ref{sec:discuss}) and a conclusion (Section \ref{sec:summary}).

\section{LBTI observations of $\beta$~Leo} \label{sec:obs}


\subsection{Instrument Setup}\label{sec:instr}

\begin{figure}[!t]
\centering
\includegraphics[height=6.5cm]{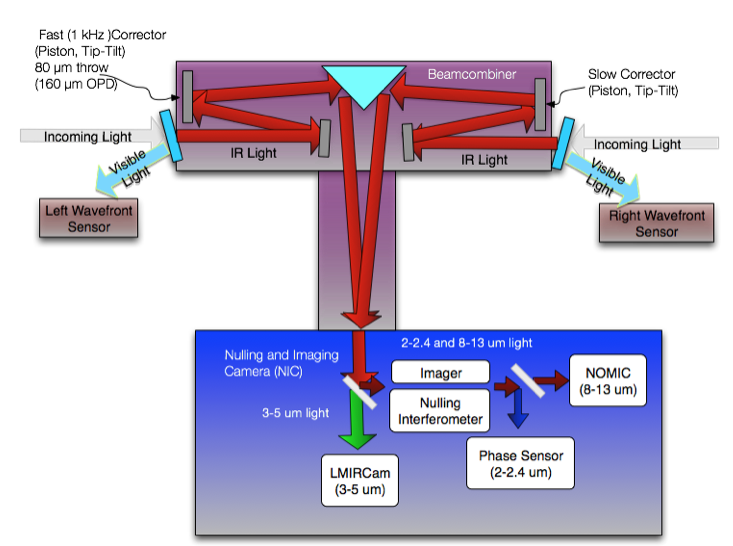}
\caption{System-level block diagram of LBTI architecture showing the optical path through the telescope, beam combiner (purple box), and the NIC cryostat (blue box). After being reflected from the LBT primaries, secondaries, and tertiaries, the visible light is reflected from the entrance window and used for wavefront sensing while the infrared light is transmitted into LBTI, where all subsequent optics are cryogenic. The beam combiner directs the light with steerable mirrors and can adjust the pathlength for interferometry. Inside the NIC cryostat, the thermal near-infrared (3-5 $\upmu$m) light is directed to LMIRCam for exoplanet imaging, the near-infrared (1.5-2.5 $\upmu$m) light is directed to the phase sensor, which measures the differential tip/tilt and phase between the two primary mirrors, and the mid-infrared (8-13 $\upmu$m) light is directed to NOMIC for nulling interferometry. Both outputs of the beam combiner are directed to the phase and tip/tilt sensor, while only the nulled output of the interferometer is reflected to the NOMIC camera with a short-pass dichroic. Note that this diagram is schematic only and does not show several additional optics.\\}
\label{fig:Comp}
\end{figure}

\begin{table*}[!t]
\begin{center}
\caption{Basic properties of $\beta$~Leo and its nulling calibrators.}\label{tab:calib}
    \begin{tabular}{c c c c c c c c c c}
        \hline
        \hline
        ID & HD & RA & DEC & Type & $m_V$ & $m_K$  & $F_{\rm \nu,N'}$ & $\theta_{\rm LD}\pm1\sigma$ &  Refs. \\
        	& 	& & & 		& 		& 			& [Jy] & [mas] & \\
        \hline
        
        $\beta$~Leo & 102647 & 11 49 03.6 & +14 34 19.4 & A3V   & 2.12 $\pm$ 0.004 & 1.91 $\pm$ 0.021 & 5.4 & 1.43 $\pm$ 0.02 & [Ge99], [Kh09] \\
        $\omicron$\,Vir & 104979 & 12 05 12.5 &  +08 43 58.7 & GIII & 4.11 $\pm$ 0.002 & 1.87 $\pm$ 0.029 & 6.1 & 1.99 $\pm$ 0.03 & [Ge99], [Kh09] \\
        25\,Com 	 & 109742 & 12 36 58.3 & +17 05 22.3 &  K5III & 5.68 $\pm$ 0.003 & 2.37 $\pm$ 0.27  & 4.2 & 1.85 $\pm$ 0.26 & [Du02], [Kh09]\\
        $\gamma$\,Com    & 108381 & 12 26 56.3 & +28 16 06.3 &  K2III & 4.34 $\pm$ 0.003 & 1.88 $\pm$ 0.090 & 6.2 & 2.12 $\pm$ 0.10 & [Du02], [Kh09] \\
        
    \hline
    \end{tabular}\\
\end{center}
    {\small References. Coordinates and spectral types from SIMBAD; V/K magnitudes and error bars from [Ge99]: \citet{Gezari1999}, [Du02]: \citet{Ducati2002}, or [Kh09]: \citet{Kharchenko2009}; N-band flux densities from \citet{Defrere2015}; Limb-darkened angular diameters and 1-$\sigma$ uncertainties computed using surface-brightness relationships \citep{Chelli2016}.}
\end{table*}

Observations of $\beta$~Leo were acquired with the Large Binocular Telescope Interferometer (LBTI) on April 24, 2013 in L' band (3.41 - 3.99 $\upmu$m) and February 8, 2015 in N' band (9.81 - 12.41 $\upmu$m). The LBTI instrument is located at the combined bent Gregorian focus of the Large Binocular Telescope (LBT) and combines the beams from the two AO-corrected 8.4-m apertures on a single detector. It has been designed primarily as a nulling interferometer as extensively described in \citet{Hinz2016} and only a quick overview is given here. The LBTI is designed to allow the use of the adaptive optics systems, and beam combination by multiple science cameras (see sketch of the instrument in Figure~\ref{fig:Comp}). Visible light from each LBT aperture is diverted via dichroics to pyramid wavefront sensors (WFS) for both the left and right apertures of the telescope. These wavefront sensors operate at 1~kHz and are clones of the pyramid wavefront sensors developed by Arcetri Observatory \citep{Bailey2010}. Each deformable mirror uses 672 actuators that routinely correct 500 Zernike modes and provide Strehl ratios as high as 80\%, 95\%, and 99\% at 1.6\,$\upmu$m, 3.8\,$\upmu$m, and 10\,$\upmu$m, respectively \citep{Bailey2014}. The bottom of Figure~\ref{fig:Comp} shows how the light is split once it enters the Nulling and Imaging Cryostat \citep[NIC,][]{Hinz2008}. The 3-5~$\upmu$m light is directed to the LMIRCam module \citep{Skrutskie2010}. The 2-2.5~$\upmu$m (NIR) and 8-13~$\upmu$m (MIR) light is reflected to the nulling interferometer, including the phase sensor \citep[PHASECAM,][]{Defrere2014} and the 8-13 $\upmu$m Nulling Optimized Mid-Infrared Camera \citep[NOMIC,][]{Hoffmann2014}. 

\begin{table}[!t]
\begin{center}
\caption{Overview of nulling observations of $\beta$~Leo, carried out on UT February 8, 2015. PA stands for parallactic angle. }\label{tab:nullobs}
    \begin{tabular}{c c c c c}
        \hline
        \hline
        Object & Time [UT] & Elevation [deg] & PA [deg] & Seeing ["] \\   
        \hline
        HD 104979	&	10:39-11:01	& 65.1-63.5 & 13.7-23.1 & 0.77-0.96 \\
        $\beta$~Leo &  11:15-11:28	& 64.4-62.4 & 41.1-44.7 & 0.74-0.83  \\
        HD 109742	& 11:38-11:56	& 70.1-67.4 & 35.3-42.7 & 0.71-0.86  \\
        $\beta$~Leo &  12:13-12:50	& 53.9-46.8 & 53.9-57.3 & 0.70-0.80  \\
        HD 108381	& 13:00-13:16	& 58.2-55.0 & 72.6-72.3 & 0.74-0.84  \\
        HD 109742	& 13:34-13:47	& 48.7-45.9 & 59.9-60.3 & 0.70-0.76  \\
    \hline
    \end{tabular}\\
    \end{center}
\end{table}

\subsection{L-band imaging observations}

$\beta$~Leo was observed with LBTI/LMIRCam at 3.8~$\upmu$m on 2013 Apr 24 UT to detect giant planets at a separation of 0.3-5 arcsec (i.e., 3-55\,au at the distance of $\beta$~Leo) as part of the LEECH survey \citep{Stone2018}. The data were collected in single-aperture direct imaging mode, using only the left side 8.4\,m aperture at the LBT, and around meridian transit to maximize field rotation. To track time-variable sky background and detector drifts, we nodded the star up and down in elevation with a throw of 4.5 arcsec every 50 frames. Each saved frame consisted of 3 co-added images with an exposure time of 0.8733 seconds. A total of 3940 frames of sufficient quality were acquired for a total on-source integration time of approximately 57 minutes. During this time, the parallactic angle changed by 61.6$\degree$ and the seeing fluctuated between 1.0 and 1.4 arcsecs.

    

\subsection{N'-band nulling observations}

$\beta$~Leo was observed with LBTI/NOMIC at 11~$\upmu$m (N' filter) on UT February 8, 2015 during nulling commissioning. The observations followed the typical sequence developed for the HOSTS survey, alternating science targets and calibrators as follows: CAL1-SCI-CAL2-SCI-CAL3-SCI-CAL2. The selected calibrator stars are listed in Table~\ref{tab:calib} and were chosen using the \texttt{SearchCal} software developed by the JMMC \citep{Bonneau2011}. In the present analysis, the third SCI pointing on $\beta$~Leo was eliminated, due to the object being too low in the sky, resulting in poor data quality. Observations were carried out for each star, by first ``locking" and optimizing the performance of the AO system for each aperture, followed by scanning and ``locking" the pathlength control loop, and tuning the pathlength setpoint to null the star at 11~$\upmu$m. A series of 1000 60~ms-long individual frames were then acquired and are together referred to as an observing block (OB). The telescope was then offset by 2.3~arcsec (keeping the star on the detector), all loops were closed, and 1000 additional frames were acquired, and then the telescope offset was removed. This cycle was repeated four times, resulting in 8 unique OBs. Measurements of the separated telescope images were then acquired, as well as a blank sky sequence. The complete sequence defines a pointing and is hence composed of several successive OBs at null, i.e.\ with the beams from both apertures coherently overlapped in phase opposition, one OB of photometric measurements with the beams separated on the detector, and one OB of background measurements with the beams nodded off the detector. Table \ref{tab:nullobs} lists the relevant parameters for each pointing of the nulling observations.


\begin{figure}[!t]
\centering
\includegraphics[height=7cm]{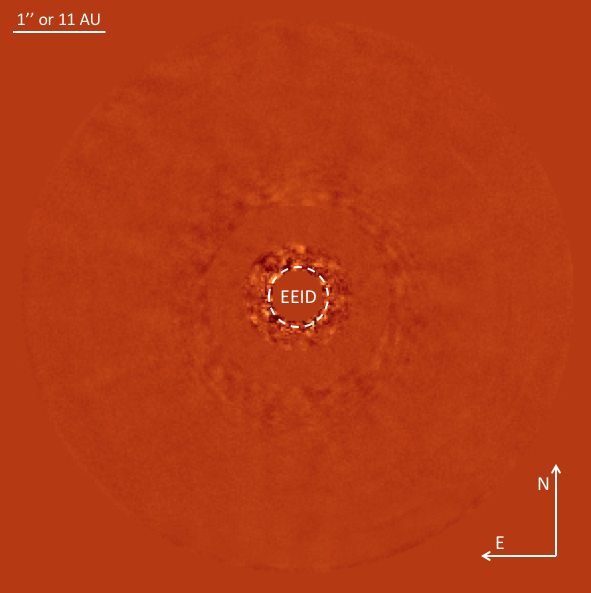}
\caption{LBTI/LMIRcam final reduced image of $\beta$~Leo at L' band after annular PCA processing with the LEECH-survey pipeline. The field of view is 6.2 arcsec and the image binned (2$\times$2 pixels). No companion is detected. The image is shown in linear intensity scale. The size of the EEID (0.335 arcsec in radius) is given by the dashed inner circle and represents the region probed by the nulling observations. The corresponding contrast curve is shown in Figure~\ref{fig:contrast}.\\}
\label{fig:img}
\end{figure}

\section{Data reduction} \label{sec:reduction}

Data reduction and analysis were carried out by the standard pipelines developed at the University of Arizona for the HOSTS and LEECH surveys as briefly described in the following sub-sections.

\subsection{Imaging L-band data}

\begin{figure}[!t]
\centering
\includegraphics[height=7.84cm]{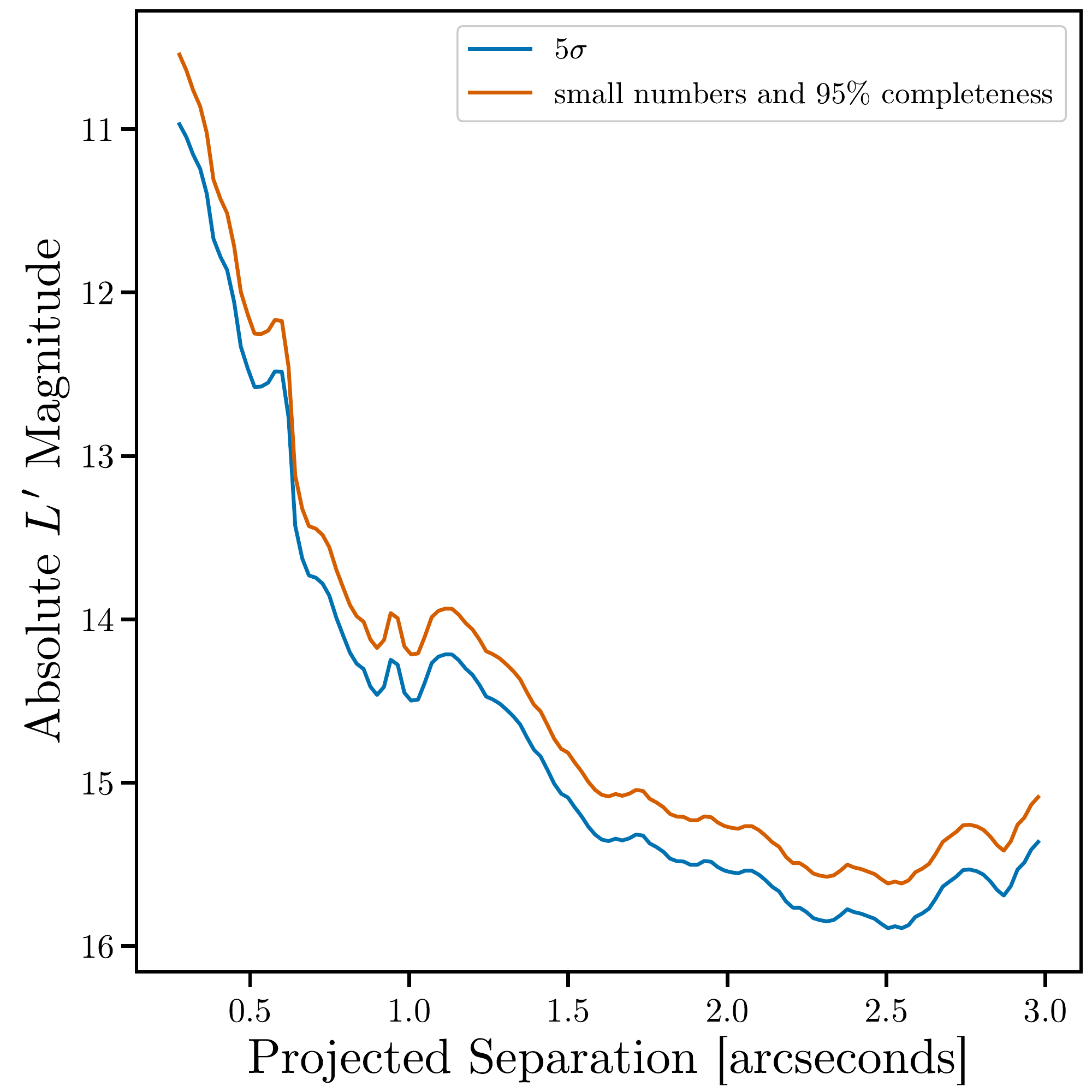}
\includegraphics[height=7.84cm]{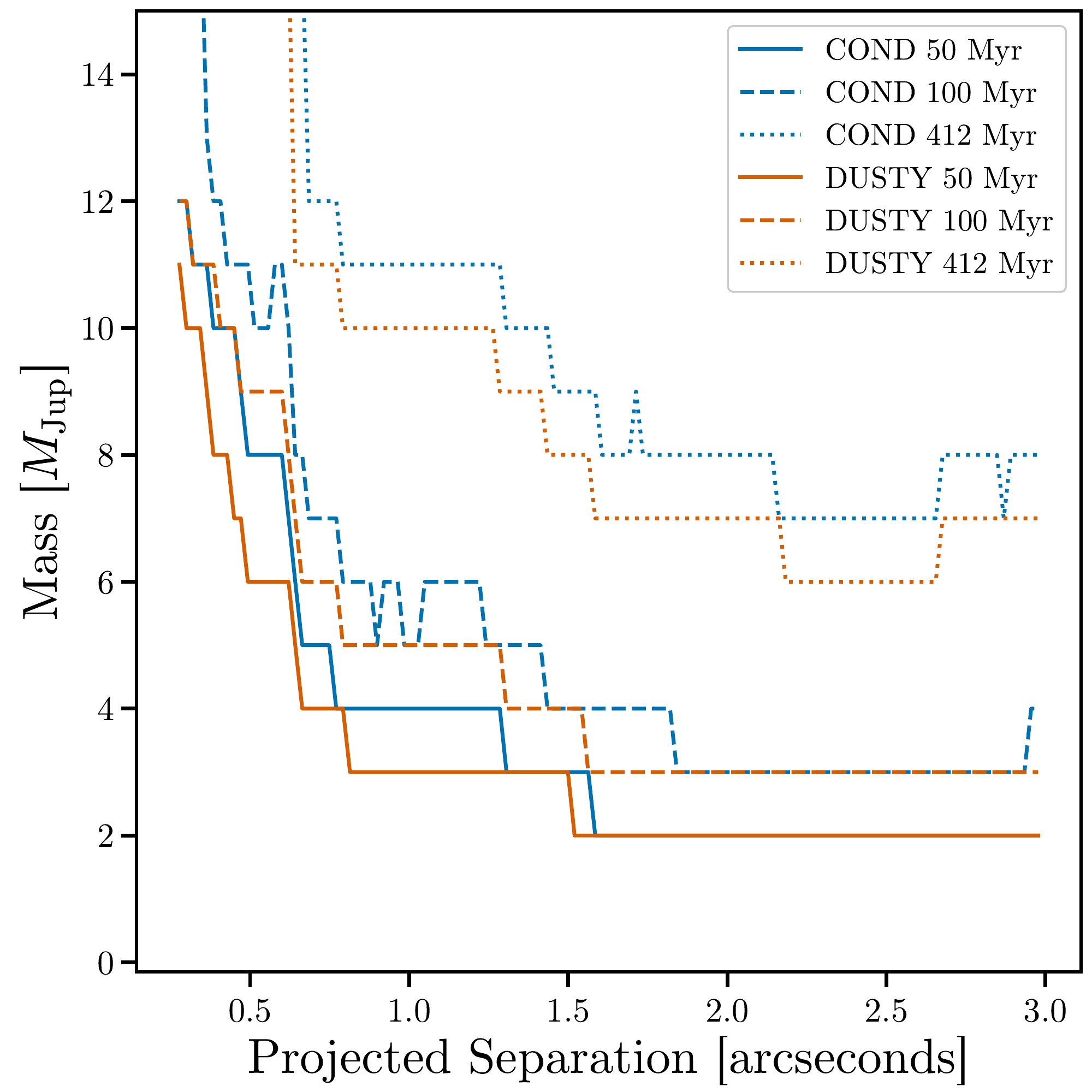}
\caption{Detection limits for giant planets around $\beta$~Leo given as 5-$\sigma$ contrast (top) and planetary mass (bottom) with respect to angular separation. In the top panel, the line labeled ``small number and 95\% completeness'' is computed for a constant number of expected false detections per radius, properly accounting for small number statistics, and ensuring 95\% completeness \citep[][]{Stone2018}. In the bottom panel, the planet mass is estimated for three different representative ages and two different evolutionary models (COND and DUSTY, see main text for more information). \\  }
\label{fig:contrast}
\end{figure}

Data were reduced using the standard LEECH-survey pipeline \citep{Stone2018}. In short, this pipeline implements the following basic image processing steps. Bad pixels are fixed by replacing their values with the median of the nearest eight good pixels. We subtract the median of each detector channel from the corresponding pixel columns to correct for bias drifts on timescales shorter than our nods. Background emission is removed from each image by subtracting the median of the 50 images in the opposite nod position taken closest in time. Each image is corrected for distortion using the dewarp coefficients reported by \citet{Maire2015}. The PSF at 3.7\,$\upmu$m has a Full Width at Half Maximum (FWHM) of 95\,mas and is over-sampled by the 10.7~mas pixels by a factor of approximately 4, so each image is binned 2x2, which has the effect of removing any residual bad pixels or cosmic ray hits.  Binned images are registered using a cross-correlation and then median combined into sets of 20 or sets with less than two degrees of rotation.  This rotation limit is chosen so that a companion at the edge of the reduced 3''$\times$3'' field of view will move by $\simeq$1 PSF width.

Our high-contrast data analysis also made use of the LEECH pipeline, which implements principal component analysis \citep[PCA,][]{Soummer2012,Amara2012,Gomez2017} to fit and remove the influence of the central star before de-rotating and stacking images. Our PCA algorithm proceeds annulus by annulus using a width of 9 pixels ($\sim$2 $\lambda/D$) to optimize the psf model, and an annulus of 1-pixel width as a subtraction region. We optimize the number of principal components at each radius by injecting fake planets and iterating until we reach the best contrast. We reduce the up and down nod positions independently and combine them using a weighted mean with weights chosen for each annulus in the image to maximize our sensitivity to artificially injected planets. Combining nods as the last step allows us to down-weight regions of poor sensitivity due to diffraction from dust near an intermediate focal plane within LMIRCam. The final image is shown in Figure~\ref{fig:img} and does not show any point-like feature. The corresponding detection limits are shown in Figure~\ref{fig:contrast} in terms of 5-$\sigma$ contrast (top) and planetary mass (bottom). They are discussed in more detail in Section~\ref{sec:results}.

\subsection{Nulling N'-band data}

\begin{figure}[!t]
\centering
\includegraphics[width=0.45\textwidth]{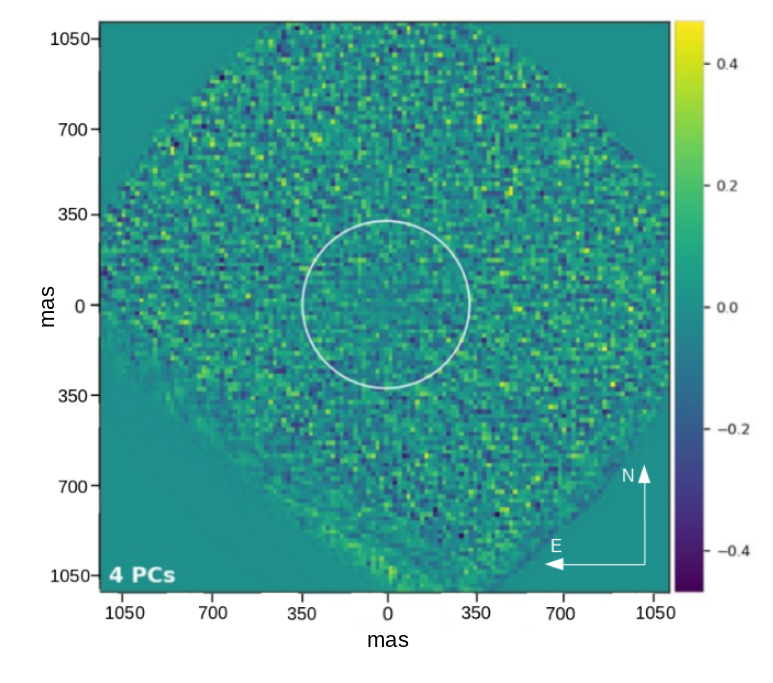}
\caption{LBTI/NOMIC final reduced image of $\beta$~Leo at N' band. The star light is removed through nulling interferometry, keeping only the frames where the null depth is below 12\%. The total integration time amounts to 920 seconds and the images were binned by 2$\times$2 pixels. The RDI processing was performed using the calibrator HD109742 and the PSF was reconstructed using the full-frame PCA method, keeping only four principal components. The Earth-Equivalent Insolation Distance (EEID) of $\beta$~Leo is represented by the solid line circle at a radius of 335~mas. The colorbar is given in ADU.\\}
\label{fig:imgN}
\end{figure}

Nulling data reduction was performed using the LBTI pipeline extensively described by \citet{Defrere2016}. In short, the raw images are corrected for bad pixels and then background subtracted using the neighboring observing blocks (OBs) taken in the opposite nods. Flux is then computed for each background-subtracted frame. In \citet{Defrere2016}, the present data set is reduced using the standard photometric aperture radius of 0.514$\lambda$/D, where $D$ is the diameter of the primary mirrors. This corresponds to the optimum size to maximize the photometric point-source SNR and is equivalent to a radius of 140\,mas (or 8 pixels) at 11\,$\upmu$m. For typical HOSTS survey stars \citep[see][]{Weinberger2015}, this is a good match to the size of the Earth-Equivalent Insolation Distance (EEID). In other words, for a solar-type star at 10\,pc distance, the 1-au EEID corresponds to a radius of 0.1 arcsec. For the more luminous, nearby stars in the sample such as $\beta$~Leo, there is significant information in the apertures of different size, which we can exploit to learn about the radial dust distribution. Hence, to measure the amount of dust in the habitable zone, flux computation has been performed for various photometric aperture sizes ranging from 2 pixels (or 36~mas) to 32 pixels (or 576 mas) in radius, which corresponds to the maximum space available on the detector to reliably compute the flux with a circular aperture. Null computation is then performed for each OB using the nulling self calibration (NSC) approach developed for the Palomar Fiber Nuller \citep[PFN,][]{Hanot2011,Mennesson2011} and adapted for the LBTI \citep{Defrere2016,Mennesson2016}. The advantage of this technique is to remove the error in the nulling setpoint between the science star and its calibrators. Finally, the instrumental transfer function is estimated by subtracting the contribution from the star (or the geometric stellar leakage) from the measured null for each OB. Limb-darkened angular diameters and 1-$\sigma$ uncertainties are computed using surface-brightness relationships \citep{Chelli2016} based on V- and K-band magnitudes. The V- and K-band magnitudes as well as limb-darkened angular diameters are listed in Table~\ref{tab:calib}. The final calibrated nulls and their corresponding error bars are shown in Table~\ref{tab:nulls} and Figure~\ref{fig:DustModel} for different radii of the photometric aperture. The excess is detected in all apertures and increases from 0.47\%$\pm$0.050\% within a radius of 1.5~au (or 140\,mas) to  1.16\% $\pm$ 0.333\% with the largest aperture (6~au or 570\,mas). As expected, the error bar is smallest for a radius  of 8 pixels (i.e., 140\,mas or 1.5~au at the distance of $\beta$~Leo), which corresponds to the optimum size to maximize the photometric SNR, and increases both toward smaller (less flux) and larger (more background noise) aperture radii.

\begin{figure}[!t]
\centering
\includegraphics[width=0.48\textwidth]{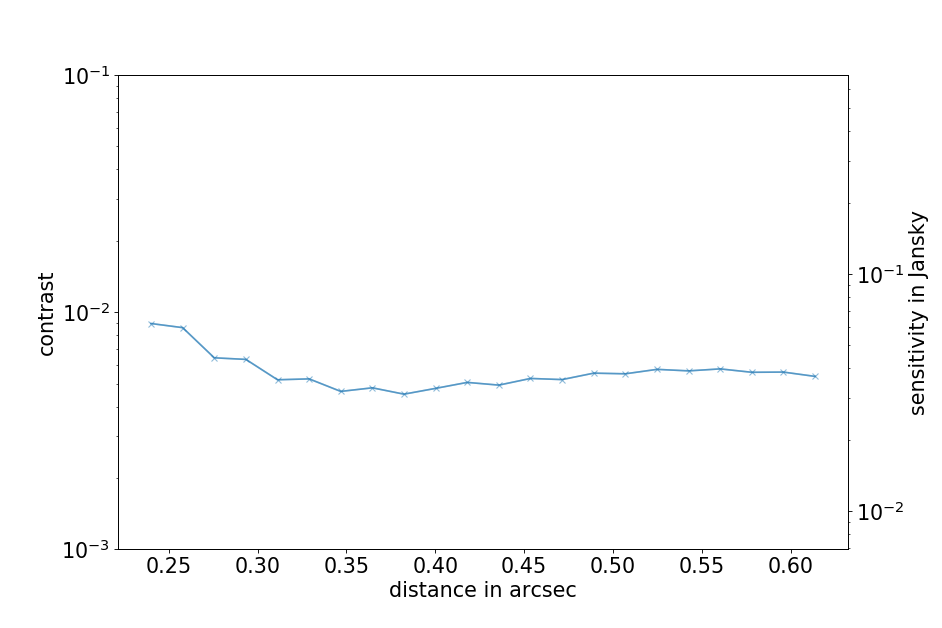}
\caption{Final 5-$\sigma$ contrast curve obtained for $\beta$~Leo at N' using the nuller as coronagraph and using RDI processing with the calibrator HD109742 as reference PSF}. Data processing and filtering were tuned to maximize the contrast at small angular separation, keeping only the frames with a null depth smaller than 12\%. The total integration time amounts to 920 seconds (or approximately 15 minutes).
\label{fig:contrastN}
\end{figure}

\begin{table}[!t]
\begin{center}
\caption{Final calibrated nulls for different aperture radii.}\label{tab:nulls}
    \begin{tabular}{c c}
        \hline
        \hline
        Aperture radius [mas] & Source null [\%] \\
        \hline
        35.7 & 0.36 $\pm$ 0.230 \\
        71.4 & 0.39 $\pm$ 0.150 \\
        143 & 0.47 $\pm$ 0.050 \\
        179 & 0.42 $\pm$ 0.054 \\
        285 & 0.54 $\pm$ 0.100 \\
        429 & 0.81 $\pm$ 0.270 \\
        571 & 1.16 $\pm$ 0.333 \\
        \hline
    \hline
    \end{tabular}\\
\end{center}
\end{table}

\begin{figure*}[!t]
\centering
\includegraphics[width=0.49\textwidth]{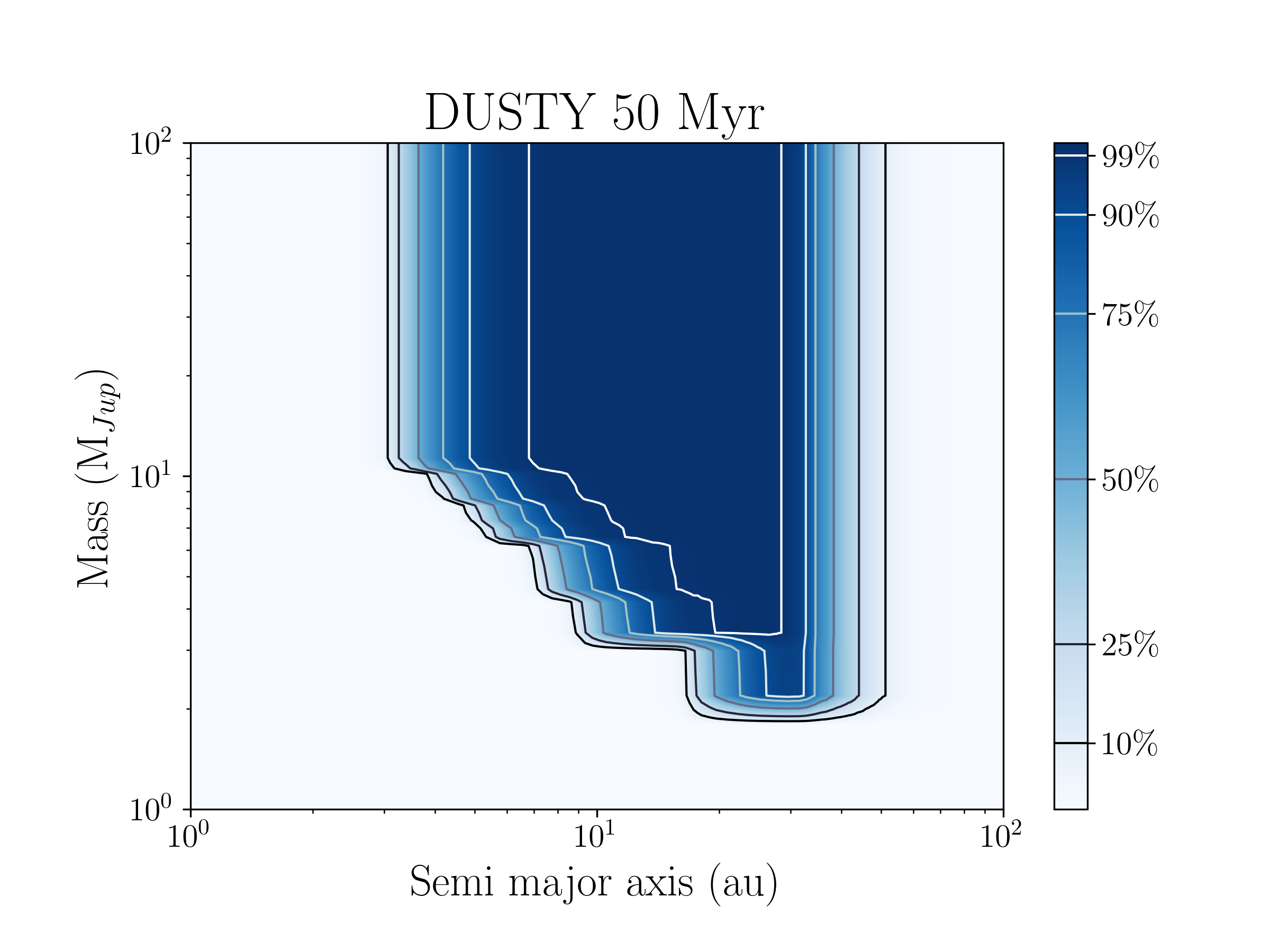}
\includegraphics[width=0.49\textwidth]{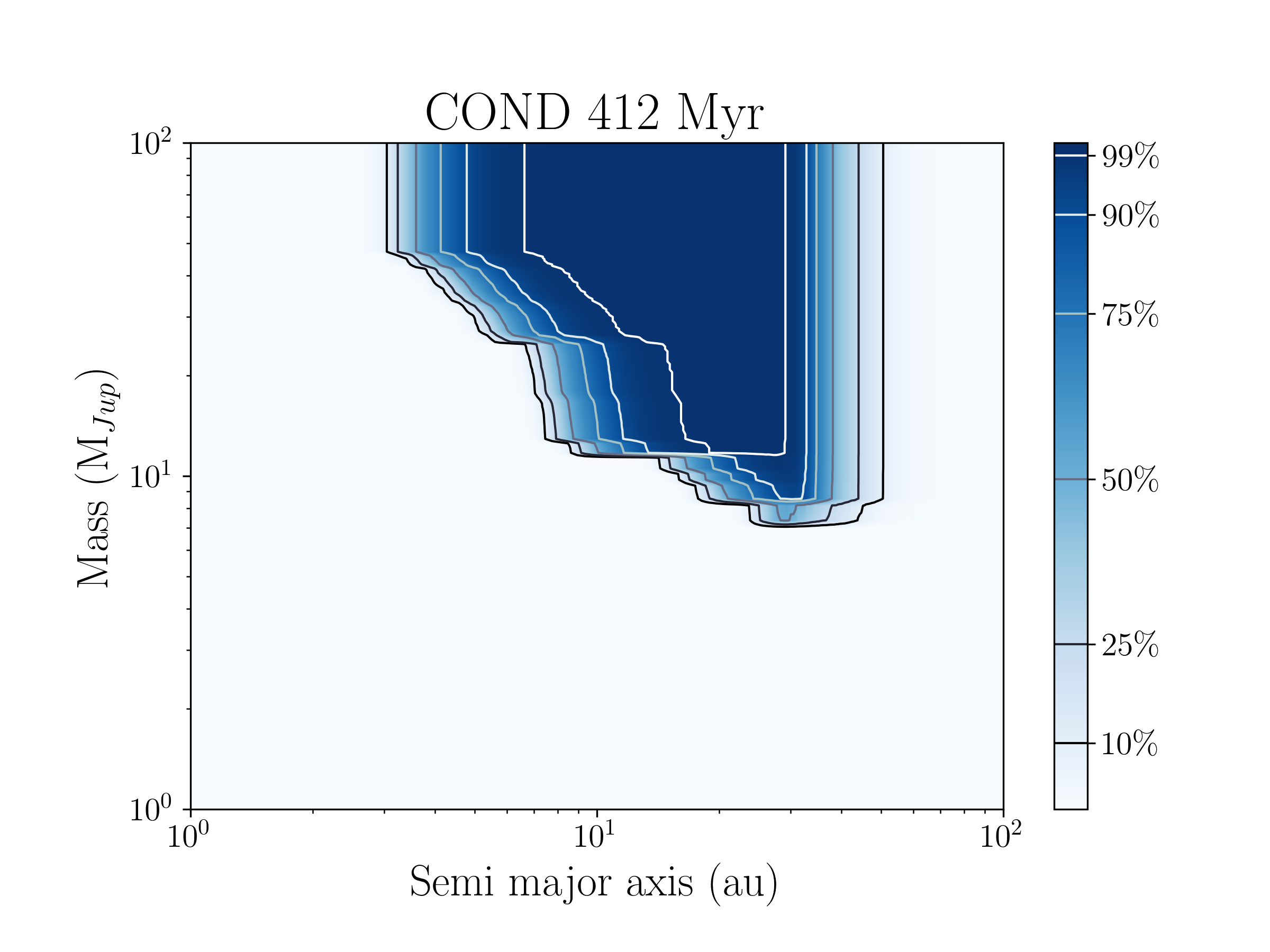}
\caption{Sensitivity maps showing the planet detection probability around $\beta$~Leo as quantified using Exo-DMC (bluescale and contours) and computed for two extreme cases: 50\,Myr with the DUSTY model (left) and 412\,Myr with the COND model (right). As illustrated in Figure~\ref{fig:contrast}, the difference in sensitivity between the two figures is primarily due to the age rather than the different evolutionary models.}
\label{fig:completenness}
\end{figure*}

In addition to the standard nulling data processing, we also applied classical direct imaging processing techniques to the background-subtracted images taken at null to look for resolved emission beyond the diffraction limit of the individual primary mirrors. Images with poor nulling performances were discarded to optimize the contrast at small angular separations while preserving sufficient frames for sensitivity purposes. After several tests, the optimal selection was found for a maximum null depth of 12\% resulting in a total integration time of 920 seconds (or approximately 15 minutes). Given the relatively small parallactic angle range of the nulling observations, we applied Reference Differential Imaging (RDI) using a single calibrator as reference (HD109742) for the PSF reconstruction using full-frame PCA implemented in the Vortex Imaging Pipeline \citep[VIP,][]{Gomez2017}. The final image is shown in Figure~\ref{fig:imgN} and the corresponding contrast curve in Figure~\ref{fig:contrastN}. No resolved emission is detected.

\section{Results and models}\label{sec:results}
\subsection{Searching for giant planets at L' band}

Figure~\ref{fig:img} does not show any particular feature in the final reduced image. To compute the detection limits corresponding to this image, we estimated the noise level as the standard deviation of the pixel intensity in concentric annuli and corrected for the self-subtraction of off-axis point sources by introducing fake companions directly in the data cube. Following the approach described in \cite{Stone2018}, the resulting 5-$\sigma$ contrast curve (see Figure~\ref{fig:contrast}, top) is used to estimate the detection limits in terms of planet mass (see Figure~\ref{fig:contrast}, bottom). We choose to use two different evolutionary models to derive two separate estimates of our sensitivity to gas-giant exoplanets for three representative ages (i.e., 50, 100, and 412\,Myr). These evolutionary models are DUSTY \citep{Baraffe2003} and COND \citep{Baraffe2003}, which represent atmospheric extremes with respect to dust and cloud opacity. DUSTY models produce atmospheres with maximal dust opacity, retaining in the photosphere all the dust and condensates that form. COND models on the other hand assume no photospheric dust opacity, but assume that dust forms and immediately precipitates below the photosphere (taking its constituent molecular species with it). As a next step, we compute the probability of detection as a function of semi-major axis and planet mass using the Exoplanet Detection Map Calculator (Exo-DMC\footnote{https://ascl.net/2010.008}). Exo-DMC is the latest rendition of the MESS code \citep[Multi-purpose Exoplanet Simulation System,][]{Bonavita2020}, a Monte Carlo tool for the statistical analysis of direct imaging survey results. In short, it combines the information on the target stars with the instrument detection limits to estimate the probability of detection of a given synthetic planet population, ultimately generating detection probability maps. For each star in the sample, Exo-DMC produces a grid of masses and physical separations of synthetic companions, then estimates the probability of detection given the provided detection limits. In the case of direct imaging observations, in order to account for the chances of each synthetic companion to be in the instrument's field of view, a set of uniformly distributed orbital parameters is generated for each point in the grid, which allows estimating the range of possible projected separations corresponding to each value of semi-major axis. The detection probability is then calculated as the fraction of orbital sets that, for a given mass, allows for the companion to be detected. In a similar fashion to its predecessors, Exo-DMC allows for a high level of flexibility in terms of possible assumptions on the synthetic planet population to be used for the determination of the detection probability. The default setup, which is the one used in this case, uses a flat distributions in log space for both the mass and semi-major axis and a Gaussian eccentricity distribution with $\mu=0$ and $\sigma= 0.3$ \citep[following the approach by][]{Hogg2010, Bonavita2013}. In addition, we use a sigma of 0.1 for the eccentricity distribution and restrict the inclination and the longitude of the node of each orbital set to make sure that all companions in the population would lie in the same orbital plane as the disk (see Section~\ref{sec:mod}). Figure~\ref{fig:completenness} shows the resulting planet detection probability computed for two extreme cases: 50\,Myr with the DUSTY model (left) and 412\,Myr with the COND model (right). For the younger age (i.e., 50\,Myr), our observations are sensitive to exoplanets down to a few Jupiter masses and located between 5\,au (or 0.5$\arcsec$) and 50\,au (or 4.5$\arcsec$) from the central star. Any planet within the sensitivity map is excluded by our LBTI imaging observations with the corresponding confidence level (see color bar).



\begin{figure*}[!t]
\centering
\includegraphics[height=6.2cm]{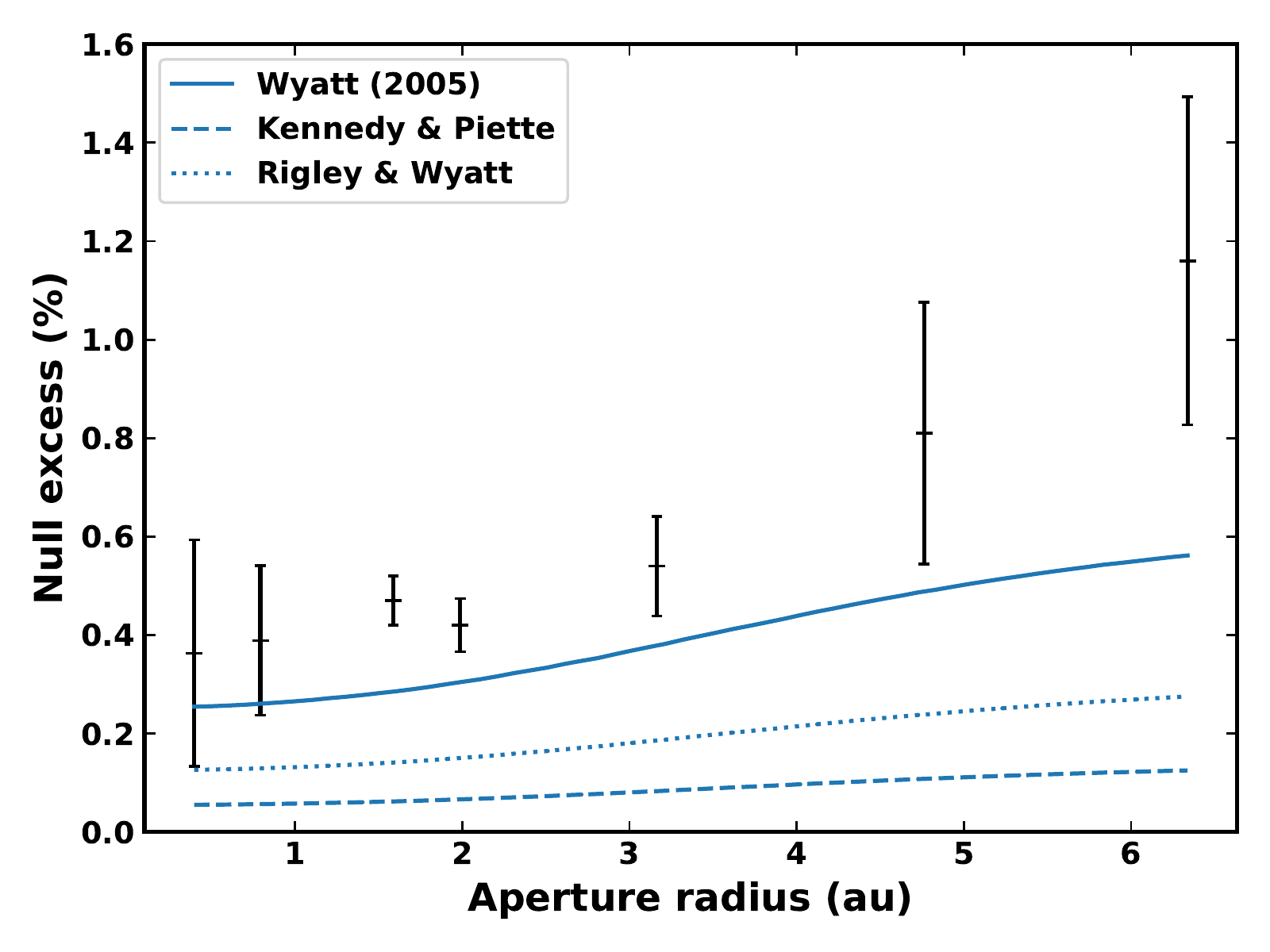}
\includegraphics[height=6.2cm]{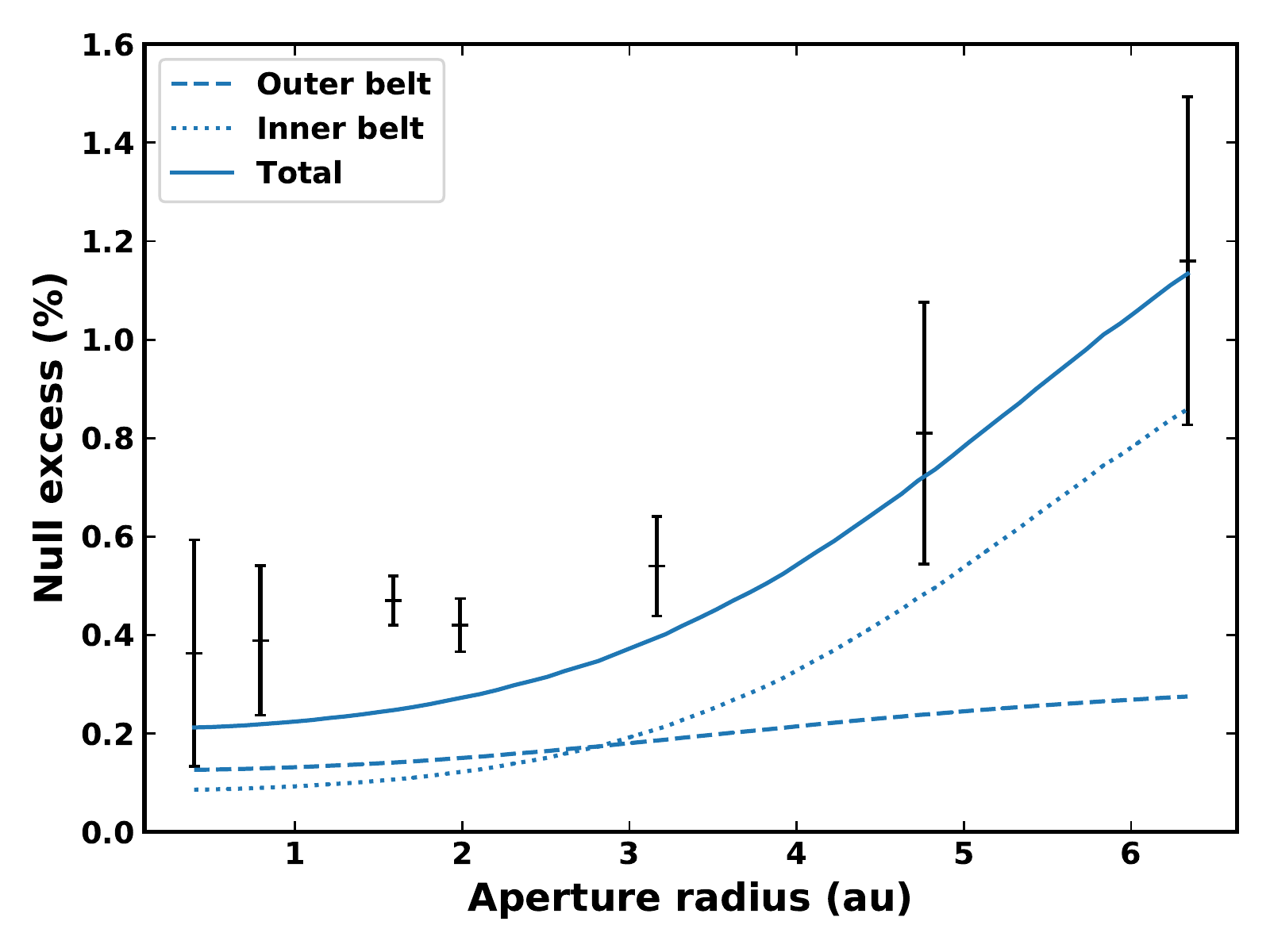}
\includegraphics[height=6.2cm]{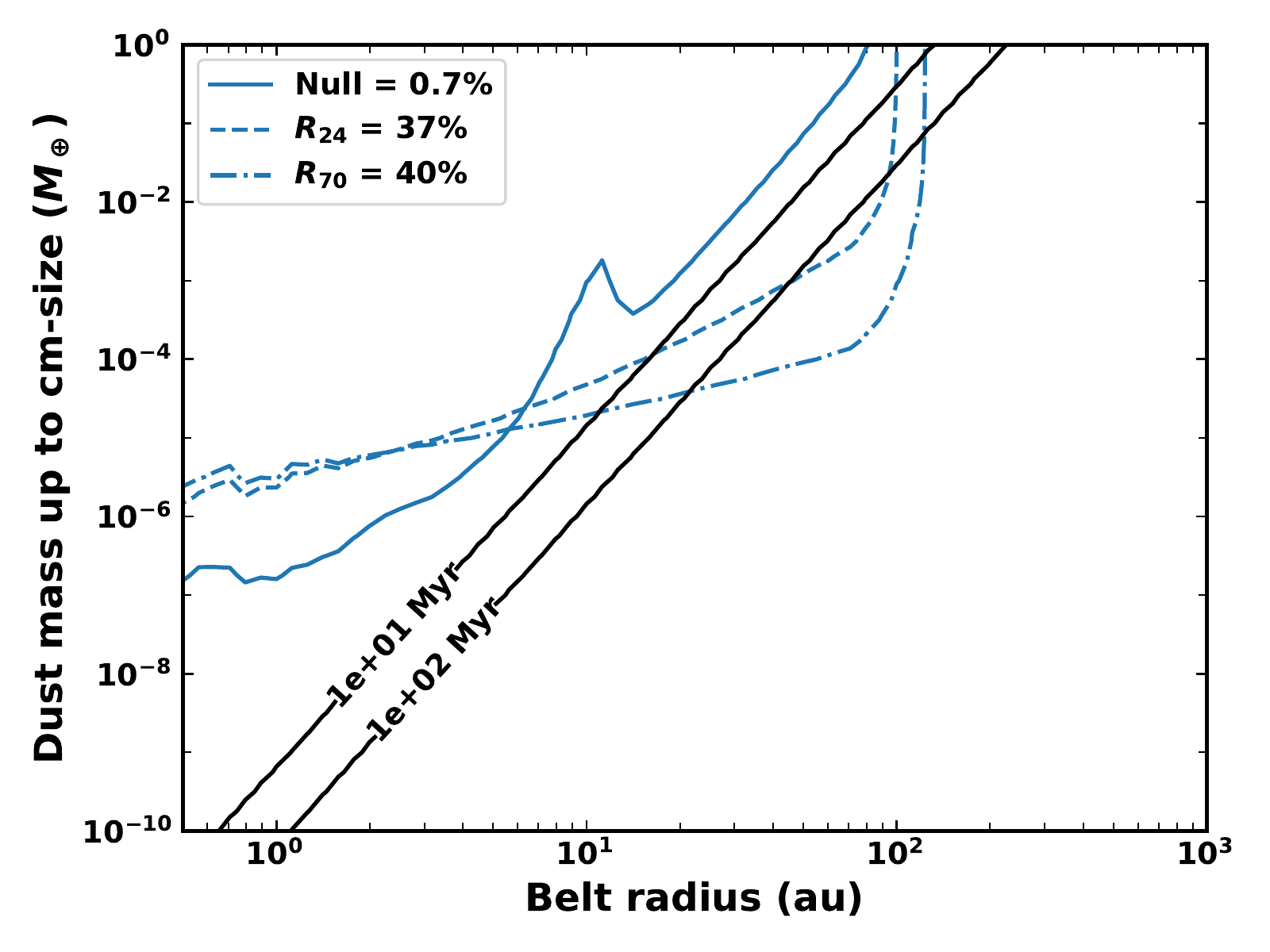}
\includegraphics[height=6.2cm]{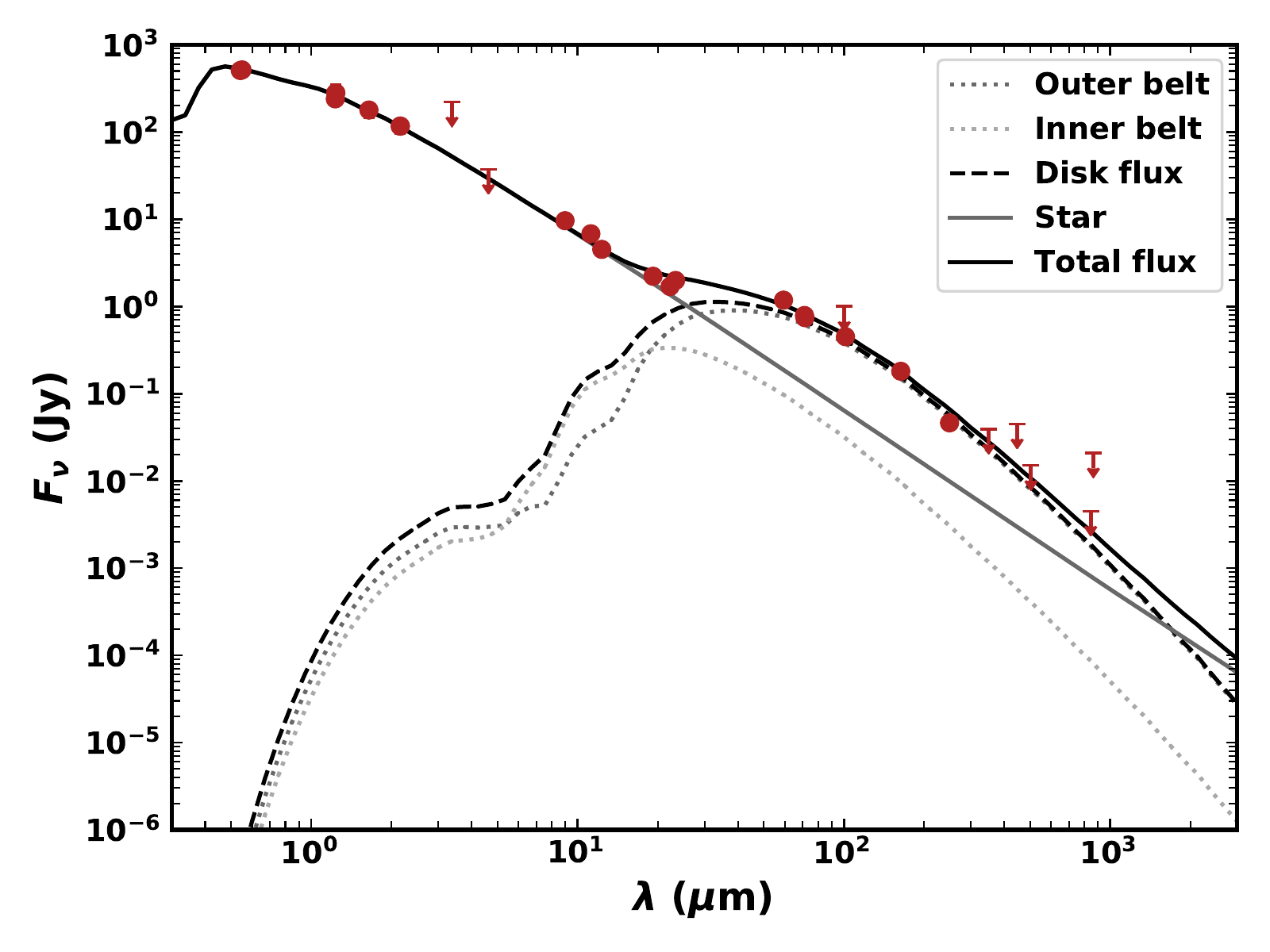}
\caption{Models invoking P-R drag to explain the null measurements to $\beta$~Leo. The top row shows null vs aperture for the observation (symbols) and three different models (lines). Top left - Analytical model using single sized black body grains \citep{Wyatt2005}, an empirical correction factor k to consider dust from a single planetesimal belt at 30\,au \citep{Kennedy2015b}, and a more comprehensive model using a realistic size distribution and optical properties \citep{Rigley2020}. Top right - more comprehensive model, in which there are two belts at 5.5\,au and 30\,au. Bottom left - Parameter space of dust mass vs belt radius used to determine parameters of the intermediate belt in the model shown Top Right. The solid black lines show the maximum dust mass for which a belt could be in steady state considering two different representative ages (10 and 100\,Myr).} Bottom right - SED of $\beta$~Leo including the spectrum of the two belt model from Top Right.
\label{fig:DustModel}
\end{figure*}





\subsection{Warm dust detected at N'-band}\label{sec:mod}


The increase in null depth with respect to the radius of the photometric aperture (see Table~\ref{tab:nulls}) is sufficient information to warrant further investigation into the structure and origin of the disk detected by the LBTI. Given the presence of an outer cold belt at $\sim$40\,au, we focus here on a P-R drag model, which is the simplest model possible to explain the observations. It can be noted that a P-R drag component is an inevitable consequence of the collisional cascade that is feeding the outer belt, so this is simply a more realistic model than not including that component. The system of interest here, $\beta$\,Leo, was among the KIN sample of stars \citep{Mennesson2014} that was used to show reasonable agreement with a P-R model \citet{Wyatt2005}. This P-R drag model solves the continuity equation for dust interior to a source belt. As the dust drifts inwards by P-R drag, it is depleted by collisions at a rate that depends on the local level of dust. The result is that the dust surface density decreases towards the star, and in most cases reaches a near-constant level close to the star. The main parameters of this model are the location and optical depth of the source region, which are set at the inner edge of the cool disk component seen by IRAS, Spitzer, and Herschel. An additional parameter $k$, introduced by \citet{Kennedy2015b}, parameterises the additional depletion in dust surface density predicted by the detailed numerical collisional model of \citet{vanLieshout2014}, relative to the assumptions of the original model of \citet{Wyatt2005}. A factor k$=$1/7 is needed to reconcile the analytic and numerical models. Note that the effect of sublimation on the numerical model of \citet{vanLieshout2014} is limited to the structure close to the star ($>$1000\,K temperatures), whereas k$=$1/7 is required to fit the profile at larger distances where collisions and P-R drag dominate. As a first step, we can compare the null excess as a function of aperture size with this P-R model. We fix the inner edge at 30\,au, and the optical depth at this location to be $1.35 \times 10^{-5}$. The inner edge was chosen based on the modelling work of \citet{Churcher2011}, and the optical depth set to ensure a good fit to the disk spectrum (assuming a 50 au-wide outer belt). A further check is that the final P-R drag + outer belt model is consistent with 70, 100, 160, and 250~$\upmu$m Herschel images \citep[modelled with the methods used in][]{Kennedy2012a,Kennedy2012b}. We find that this model can reproduce the far-IR images well, and derive updated parameters for the outer disk geometry of position angle of 142$\degree$ and inclination of 45$\degree$ (with an uncertainty of 10$\degree$). To model the LBTI observations, we generated images of the P-R drag component, and computed aperture-corrected null excess values using the LBTI transmission pattern \citep{Kennedy2015}, the 312~mas FWHM PSF, and the N-band stellar flux density of 5.4\,Jy. As shown in Figure~\ref{fig:DustModel} (top left), both the original analytic model of \citet{Wyatt2005} and the modified model of \citet{Kennedy2015b} under-predict the amount of flux measured by the LBTI. While this suggests that a P-R drag model may not be able to explain the observed dust, strong conclusions could not be reached from this model since it only considers a single grain size and black body dust.

To consider a realistic grain size distribution as well as realistic optical properties, we used in a second step the model of \citet{Rigley2020}. This new model considers not only the spatial density profile as in the models of \citet{Wyatt2005} and \citet{Kennedy2015}, but also how the size distribution varies with distance from the star due to collisions and P-R drag as in the model of \citet{vanLieshout2014}. The optical properties that are assumed for the grains can have a significant effect on predictions for the amount of flux they produce. Optical properties of the grains are calculated using the same method as \citet{Wyatt2002}, with compositions from the core-mantle model of \citet{Li1997}, which assumes a silicate (amorphous olivine) core and organic refractory mantle, and has three free parameters. A range of compositions was used, with silicate volume fractions varying from 0 to 1, porosities from 0 to 0.95, and volume fraction of water ice in the gaps from 0 to 1. The grain size distribution in the model is then calculated self-consistently, both in the planetesimal belt and the region interior to it, from the competition between collisions and P-R drag, with a cut-off due to radiation pressure at the blow-out limit. Considering a belt with an inner edge at 30~au, the best-fit composition was found by minimising chi-squared across a grid of compositions and simultaneously fitting to the measured nulls and spectral energy distribution (SED), using the photometry and stellar model from shown in Figure~\ref{fig:DustModel}. The photometry is largely the same as used in \citet{Matthews2010} and \citet{Churcher2011}, with the addition of WISE \citep{Wright2010}, AKARI IRC \citep{Ishihara2010}, and SCUBA2  measurements \citep{Holland2017}. The stellar component is fit using the method described in \citet{Yelverton2019,Yelverton2020}, which is subtracted to yield the disk fluxes fitted by the model. The best fit is obtained for a composition with 75\% silicate and 25\% organic grains. While the flux predicted by the new P-R drag model is approximately two times higher than the model with k$=$1/7 (see dotted line in top left panel of Figure~\ref{fig:DustModel}), it is still two to four times too low to explain the LBTI nulling data.

\begin{figure}[!t]
\centering
\includegraphics[width=0.47\textwidth]{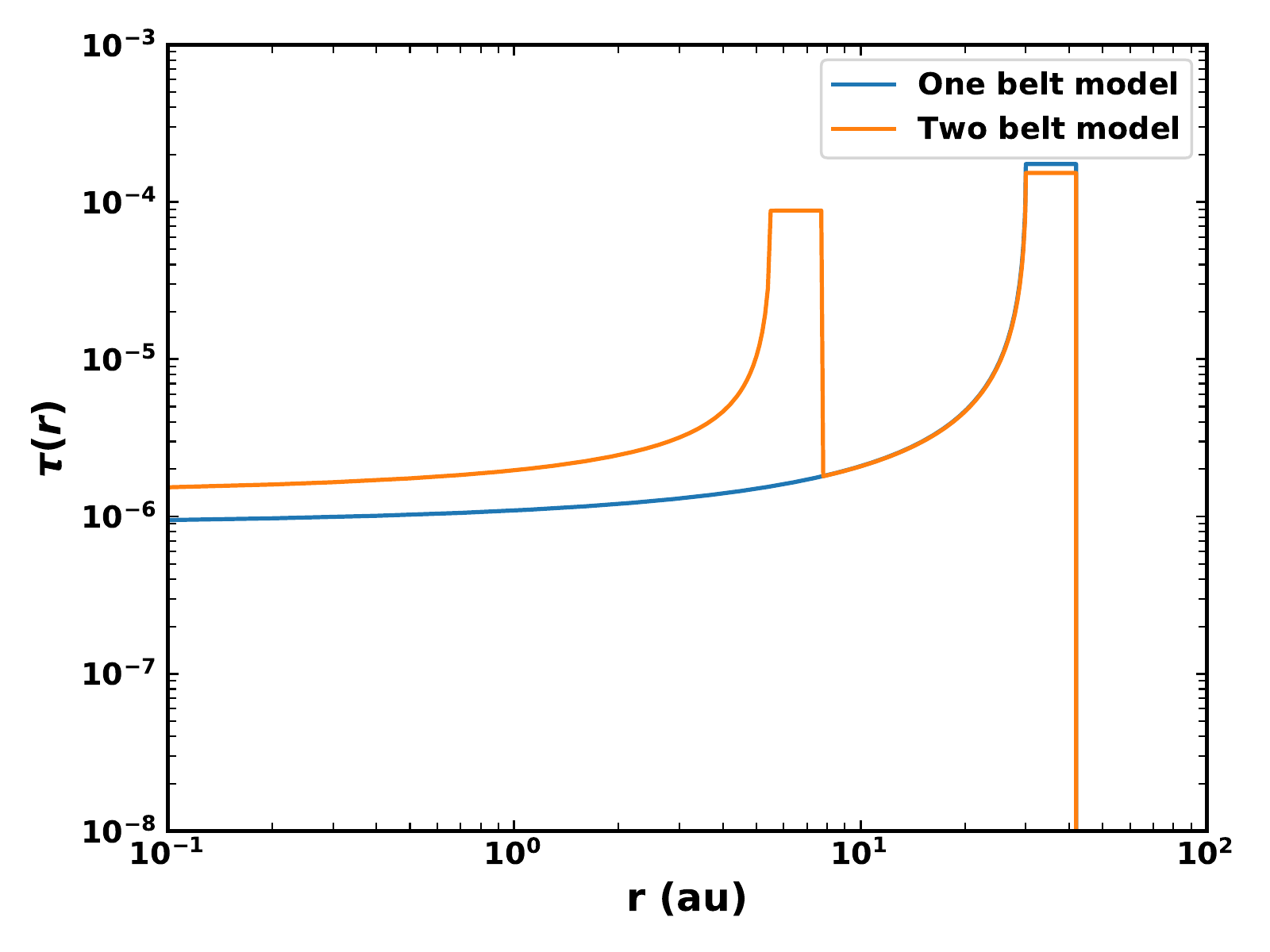}
\caption{Radial profile of the surface density of grains for the one-belt (blue line) and two-belt (orange) models. The two-belt model shows an increased level of P-R drag dust, which is required to fit our LBTI nulling data (see Figure~\ref{fig:DustModel}, top right).}
\label{fig:taur}
\end{figure}

One possibility to reconcile the model with the data is to assume that there is an additional intermediate, warm belt, as suggested by \citet{Churcher2011}, which would have been unresolved by Herschel and Spitzer. The three-component model of \citet{Churcher2011} had a hot component at 2~au, warm dust at 9~au, and cold dust from 30 to 70~au. As a toy model, we superposed the emission of the known outer belt with a second belt interior to it, assuming that both belts have the same composition. The mass of the outer belt is well constrained by the SED, as it dominates the far-infrared emission. We then use the approach of Figure~15 of \citet{Rigley2020} to determine the parameters of the inner belt that would satisfy the observational constraints. These constraints are the observed 24-$\upmu$m excess of 37\%, a 40\% excess at 70~$\upmu$m, and an 11-$\upmu$m null of 0.5 per cent within the conservative aperture. The constraints converge on a belt with an inner radius of a few au and mass $\sim10^{-5}~\mathrm{M}_{\oplus}$ (see where the blue lines converge on Figure~\ref{fig:DustModel}, bottom left). These parameters along with the composition and outer belt mass were refined with a combined chi-squared fit to both the SED and null. This gave that the best composition was the same as that used for the one belt model (since the chi-squared is dominated by the contribution from the SED). {The parameters of the inner belt, which converge on a radius of 5.5~au and mass $1.5\times\sim10^{-5}~\mathrm{M}_{\oplus}$, are primarily determined by the mid-infrared emission: the location of the belt has to be optimised so that it produces enough 11 micron flux, however it is difficult to avoid producing too much 24 micron flux with the additional warm belt. This means that it is difficult to get a perfect fit to the radial null profile given the observational constraints.}
Figure~\ref{fig:DustModel} shows the corresponding SED (bottom right) and predicted null for this two belt model (top right). The radial profile of the surface density of grains is given in Figure~\ref{fig:taur}, showing the planetesimal belts and the increase in the P-R drag dust resulting from superposing the two belts. Combining the emission of the two belts agrees well with the null measurements for the larger aperture radii, while fitting the SED. While this approach treated dust created in the outer and inner belts separately, and so ignored collisions between these populations, and still under-predicts the null for small apertures, this shows how in principle the observations are consistent with the observed null having its origin, at least partially, in an inner belt. More LBTI observations at different wavelengths and covering a wider range of paralactic angles are required to better constrain the disk geometry and to look for possible asymmetric disk structures.

\section{Discussion} \label{sec:discuss}

While studies of close-in RV planets do not find a significant correlation between the frequency and properties of debris disks and the presence of close-in planets \citep[e.g.,][]{MoroMartin2007,Bryden2009,Yelverton2020}, recent work based on a sample of 130 debris disk single stars and 277 stars that do not show infrared excesses suggest that wide-separation giant planets may be more frequent around the stars with debris disks \citep[i.e., 6.27\% compared to 0.73\% for the control sample,][]{Meshkat2017}. Even though $\beta$~Leo has an outer dust belt, the non-detection of giant planets in our L'-band data is not a surprise given the relatively low occurrence rate of giant exoplanets at these separations. This is consistent with the P-R drag model of the dust, which implies that planets more than a few Saturn masses do not reside beyond $\sim$5~au and interior to the outer belt, as these would otherwise accrete or eject the dust before it reaches the inner regions \citep{Bonsor2018}. One question however is whether an inner belt at 5.5~au with this level of dust production can be a steady state phenomenon. To address this question, we use the model of \citet{Wyatt2007} and compute the maximum dust mass for which a belt could be in steady state at a given age. These are represented by the black solid lines in Figure~\ref{fig:DustModel} (bottom left), together with the dust mass estimated from our LBTI nulling resolved observations (see solid blue line) and Spitzer photometric constraints (see dashed blue lines). As described in the previous section, the best-fit model of the inner belt has a radius of $\sim$5.5~au and a mass of $\sim1.5 \times 10^{-5}~\mathrm{M}_{\oplus}$, near the intersection of the LBTI and Spitzer lines. Based on $\beta$~Leo's age, we can conclude that the proposed inner belt is too massive to be in steady state. One possibility is that the belt is a relatively recent phenomenon, e.g., having been created in the recent break-up of a very large asteroid. Alternatively the belt could be continually replenished by comets scattered in from the outer belt, similar to the way comets replenish the dust in the zodiacal cloud \citep{Nesvorny2010}. In this case, planets would be required to scatter planetesimals onto comet-like orbits, though they may have masses far below our detection limits \citep{Bonsor2012,Marino2018}.

\begin{figure}[!t]
\centering
\includegraphics[width=0.47\textwidth]{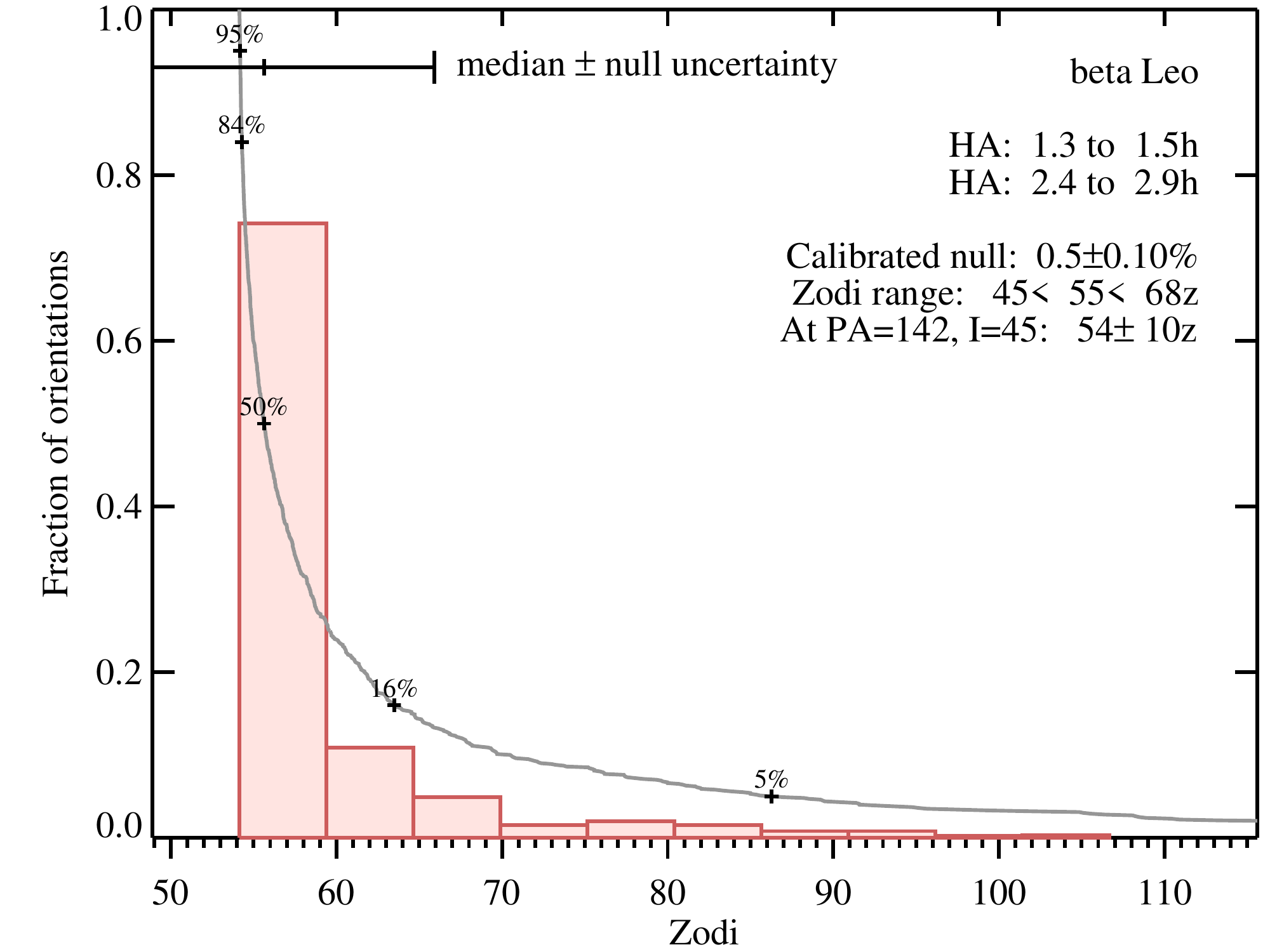}
\caption{Distribution of zodi levels over all possible disk orientations. The fraction of orientations at each zodi level is shown by the histogram, while the cumulative distribution is shown as the solid line. The component of the uncertainty that arises from the null excess measurement is given by the error bar about the median at the top left. The legend shows the hour-angle ranges, the null excess measurement, the 1$\sigma$ range of zodi levels for all possible disk orientations, and the zodi level if the disk were at the position angle and inclination of the outer disk. \\}
\label{fig:zlim}
\end{figure}

Regarding the amount of dust at radial distances where habitable planets might lie, we applied the ``standard'' HOSTS disk model, as described by \citet{Kennedy2015}, to the null excess measured with an aperture of 3~au (or 286\,mas at the distance of $\beta$~Leo). The purpose of this model is to provide a standardized set of results for HOSTS targets. The result of applying the standard model is shown in Figure \ref{fig:zlim}. This figure shows the distribution of ``zodi" levels that are allowed when the disk model is randomly distributed over all orientations, and that the range of zodi levels when the null excess uncertainty is also included is from 45 to 68 zodis. This range is dominated by the uncertainty in the null excess. If the zodiacal emission is in a disk with the same geometry as the outer disk, the zodi level would be 50$\pm$10 zodis. This high level of dust in the habitable zone of $\beta$~Leo makes it clear that it is not a good target for a future exo-Earth imaging instrument \citep[see discussions in e.g.,][]{Roberge2012,Defrere2010,Stark2015}.




\section{Summary} \label{sec:summary}

The LBT Interferometer was used to characterize the brightness and spatial extension of the warm dust belt around $\beta$~Leo, as measured at 11~$\upmu$m wavelength. The excess is measured at 0.47\%$\pm$0.05\% at 1.5~au, and rises to 0.81\%$\pm$0.26\% at 4.5~AU, outside the habitable zone region for $\beta$ Leo.  This level of dust amounts to approximately 50 times the dust from the solar system zodiacal cloud (50 $\pm$ 10 zodis) assuming the same orientation as the outer belt. Based on 70, 100, 160, and 250~$\upmu$m Herschel images, we derive updated values for the disk position angle of 142$\degree$ and inclination 45$\degree$. Models of the cold dust previously detected by Spitzer and Herschel, when combined with an evolution determined by P-R drag, under-predict the amount of dust detected in the habitable zone of $\beta$~Leo, and require an additional warm belt at approximately 5.5~au. Based on $\beta$~Leo's age, we find that this inner belt is too massive to be in steady state. This suggests that the belt is a relatively recent phenomenon, e.g., having been created in the recent break-up of a very large asteroid. Alternatively the belt could be continually replenished by comets scattered in from the outer belt by giant planets, similar to the way comets replenish the dust in the zodiacal cloud. To address this question, we present constraints from LBTI imaging at 3.8 $\upmu$m wavelength for giant planets. Assuming an age of 50\,Myr, the observations constrain any planet in the system between approximately 5\,au to 50\,au to be less than a few Jupiter masses. While this is consistent with the dust model presented in this study, these detection limits are not sufficient to distinguish between the two proposed scenarios and more sensitive observations with JWST are required to shed light on this system. These observations provide the first example of observations from the HOSTS survey to characterize typical zodiacal dust brightness levels around nearby stars. 



\begin{acknowledgements}
The LBT is an international collaboration among institutions in the United States, Italy and Germany. LBT Corporation partners are: The University of Arizona on behalf of the Arizona university system; Istituto Nazionale di Astrofisica, Italy; LBT Beteiligungsgesellschaft, Germany, representing the Max-Planck Society, the Astrophysical Institute Potsdam, and Heidelberg University; The Ohio State University, and The Research Corporation, on behalf of The University of Notre Dame, University of Minnesota and University of Virginia. LBTI is funded by a NASA grant in support of the Exoplanet Exploration Program. This research has made use of the Jean-Marie Mariotti Center \texttt{SearchCal} service \footnote{Available at http://www.jmmc.fr/searchcal} co-developped by LAGRANGE and IPAG, and of CDS Astronomical Databases SIMBAD and VIZIER\footnote{Available at http://cdsweb.u-strasbg.fr/}. This work was supported by the European Union through ERC grants number 866070 (DD) and 279973 (MCW). DD and OA acknowledge the support of the Belgian National Funds for Scientific Research (FNRS). GMK is supported by the Royal Society as a Royal Society University Research Fellow. M.B. acknowledges funding by the UK Science and Technology Facilities Council (STFC) grant no. ST/M001229/1. 
\end{acknowledgements}

\newpage
\bibliography{betaLeo.bib}
\bibliographystyle{apj}



\end{document}